\documentclass[sigplan,10pt,review=false,anonymous=false]{acmart}
\renewcommand\footnotetextcopyrightpermission[1]{}
\newcommand{\solution}{PlexRL }
\AtBeginDocument{%
  }

\setcopyright{none}
\copyrightyear{2026}
\acmYear{2026}
\usepackage{amsmath}
\usepackage[ruled,linesnumbered]{algorithm2e}

\begin{document}

\title{PlexRL: Cluster-Level Orchestration of Serviceized LLM Execution for RLVR}

\author{Yiqi Zhang}
\authornote{Work done during an internship at Shanghai Qiji Zhifeng Co., Ltd.}
\email{yiqi.zhang@u.nus.edu}
\affiliation{%
  \institution{National University of Singapore}
  \country{Singapore}
}

\author{Fangzheng Jiao}
\email{jiaofz@buaa.edu.cn}
\affiliation{%
  \institution{Beihang University}
  \country{China}
}

\author{Tian Tang}
\email{tangtian@qijizhifeng.com}
\affiliation{%
  \institution{Shanghai Qiji Zhifeng Co., Ltd.}
  \country{China}
}

\author{Boyu Tian}
\email{tianboyu@qijizhifeng.com}
\affiliation{%
  \institution{Shanghai Qiji Zhifeng Co., Ltd.}
  \country{China}
}

\author{Hangyu Wang}
\email{wanghangyu@qijizhifeng.com}
\affiliation{%
  \institution{Shanghai Qiji Zhifeng Co., Ltd.}
  \country{China}
}

\author{Qiaoling Chen}
\authornotemark[1]
\email{qiaoling.chen@ntu.edu.sg}
\affiliation{%
  \institution{Nanyang Technological University}
  \country{Singapore}
}

\author{Guoteng Wang}
\email{wangguoteng@qijizhifeng.com}
\affiliation{%
  \institution{Shanghai Qiji Zhifeng Co., Ltd.}
  \country{China}
}

\author{Zhen Jiang}
\email{jiangzhen@qijizhifeng.com}
\affiliation{%
  \institution{Shanghai Qiji Zhifeng Co., Ltd.}
  \country{China}
}

\author{Peng Sun}
\email{sunpeng@qijizhifeng.com}
\affiliation{%
  \institution{Shanghai Qiji Zhifeng Co., Ltd.}
  \country{China}
}

\author{Ping Zhang}
\email{zhangping@infrawaves.com}
\affiliation{%
  \institution{Infrawaves}
  \country{China}
}

\author{Xiaohe Hu}
\email{huxiaohe@infrawaves.com}
\affiliation{%
  \institution{Infrawaves}
  \country{China}
}

\author{Ziming Liu}
\authornotemark[1]
\email{liuziming@comp.nus.edu.sg}
\affiliation{%
  \institution{National University of Singapore}
  \country{Singapore}
}

\author{Menghao Zhang}
\email{zhangmenghao0503@gmail.com}
\affiliation{%
  \institution{Beihang University}
  \country{China}
}

\author{Yanmin Jia}
\email{jiayanmin@infrawaves.com}
\affiliation{%
  \institution{Infrawaves}
  \country{China}
}

\author{Yang You}
\email{dcsyouy@nus.edu.sg}
\affiliation{%
  \institution{National University of Singapore}
  \country{Singapore}
}

\author{Siyuan Feng}
\email{syfeng@sii.edu.cn}
\affiliation{%
  \institution{Shanghai Innovation Institute}
  \country{China}
}
\affiliation{%
  \institution{Shanghai Qiji Zhifeng Co., Ltd.}
  \country{China}
}

\renewcommand{\shortauthors}{Zhang et al.}

\begin{abstract}
Reinforcement learning with verifiable rewards (RLVR) has recently unlocked strong reasoning capabilities in large language models (LLMs), triggering rapid exploration of new algorithms and data. However, RLVR training is notoriously inefficient: long-tailed rollouts, tool-induced stalls, and asymmetric resource requirements between rollout and training introduce substantial idle time that cannot be eliminated by job-local optimizations such as synchronous pipelining, asynchronous rollout, or colocated execution.

We argue that this inefficiency is structural. While idle gaps are unavoidable within individual RLVR jobs, they are largely anti-correlated across jobs and therefore exploitable at the cluster level. Leveraging this observation, we present \solution, a cluster-level runtime for multiplexing unified LLM services across RLVR jobs. By centrally managing model placement, state transitions, and function-level scheduling under strict affinity constraints, \solution time-slices LLM execution across jobs to fill otherwise idle periods without expensive model migration.
Our implementation and evaluations demonstrate that \solution significantly improves effective cluster capacity and reduces user GPU hour cost by maximum \textbf{37.58\%} while preserving algorithmic flexibility and introducing minimal per-job overhead. 
\end{abstract}




\settopmatter{printfolios=true,printacmref=false}\maketitle
\pagestyle{plain}

\section{Introduction}

Recent advances in large language models (LLMs) have renewed interest in reinforcement learning beyond supervised fine-tuning. DeepSeek-R1 \cite{deepseekai2025deepseekr1incentivizingreasoningcapability} showed that pretrained LLMs contain substantial latent reasoning ability that can be unlocked through reinforcement learning with verifiable rewards (RLVR), yielding strong gains on mathematical, logical, and planning tasks with comparatively little task-specific supervision \cite{muennighoff2025s1simpletesttimescaling}. This success has triggered rapid exploration of new RLVR algorithms, objectives, and datasets for reasoning-centric LLMs \cite{yue2025vapoefficientreliablereinforcement,yu2025dapoopensourcellmreinforcement,xie2025logicrlunleashingllmreasoning,kimiteam2025kimik15scalingreinforcement,muennighoff2025s1simpletesttimescaling}. At the same time, both academia and industry are increasingly shifting toward \emph{agentic} LLM systems that interact with external tools, environments, and executors; by combining language modeling with tool use, code execution, and environment feedback, these systems can solve tasks that extend far beyond static text generation \cite{gou2024tora,yang2024qwen25mathtechnicalreportmathematical}. Despite their apparent diversity, these workloads share a common training backbone: models repeatedly generate trajectories, receive delayed or verifiable rewards, and improve through RL-style updates.

This trend has exposed a growing mismatch between modern RLVR workloads and existing RL training frameworks. Contemporary RLVR pipelines are increasingly heterogeneous: researchers combine different rollout strategies, reward models, verifier components, additional model roles, speculative branches, and tool-augmented interaction loops \cite{lu2025onpolicydistillation,liu2026specrlacceleratingonpolicyreinforcement,jin2025searchr1trainingllmsreason,he2026searchr2enhancingsearchintegratedreasoning}. Yet many current systems still assume relatively fixed training structures, model roles, and execution orders. Flexible libraries such as OpenRLHF and TRL \cite{hu2025openrlhfeasytousescalablehighperformance,vonwerra2022trl} make it easy to prototype new algorithms, but often leave substantial performance on the table. Conversely, highly optimized systems such as VeRL and NeMo-Aligner \cite{sheng2025hybridflow,shen2024nemo} deliver strong efficiency in their target settings through tightly integrated execution pipelines and backend-specific optimizations. A side effect of this specialization is that framework assumptions about rollout--training structure, model placement, and execution order become increasingly difficult to escape. As RLVR exploration broadens, algorithm developers are forced either to sacrifice performance for flexibility, or to perform invasive framework surgery whenever a new workflow falls outside the assumed execution template.

At the same time, above frameworks suffer from severe cluster-level inefficiency. These workloads are dominated by multi-stage execution patterns with very different performance characteristics: rollout consists of autoregressive decoding, often interleaved with long-tailed tool calls or environment interaction, while training requires repeated forward and backward passes over large actor-critic style models. As model sizes grow into the hundreds of billions of parameters, even modest inefficiencies translate directly into prohibitive infrastructure cost. In practice, RLVR workloads frequently exhibit poor accelerator utilization: split deployments waste resources through phase alternation, colocated deployments oversize rollout with training-driven parallelism, and asynchronous pipelines can only partially hide rollout--training mismatch while introducing staleness trade-offs. As a result, large fractions of reserved accelerator capacity remain idle even when clusters are nominally full.

These two problems share a common root cause: existing systems bind algorithm control to job-local model execution. This coupling not only hard-wires RL pipelines to fixed deployment topologies and backend assumptions, but also hides fine-grained execution dynamics from the cluster behind coarse job reservations. As a result, the system is simultaneously too rigid for rapidly evolving RLVR workflows and too opaque to reclaim idle capacity across jobs. 

Fortunately, modern LLM workloads also admit a cleaner systems abstraction. Although RLVR pipelines are heterogeneous in control flow, they are comparatively homogeneous in how they invoke model execution. Across reasoning and agentic workloads, model-side computation largely reduces to a small set of common primitives: token generation, forward evaluation, backward computation, optimizer step, and weight synchronization. Moreover, contemporary model families are increasingly concentrated around a small number of architectural patterns, such as Transformers and related sequence models \cite{vaswani2017attention,gu2024mamba,gu2022s4}. This combination---\emph{heterogeneous pipelines, homogeneous model calls}---suggests that algorithm control should be decoupled from the execution substrate. Exposing training and inference through unified service interfaces not only gives algorithm developers a hassle-free framework for composing new RL workflows, but also creates the cluster-visible execution layer required for global orchestration.

Building on this observation, we present \solution, a cluster-level multiplexing system for unified LLM services tailored to RLVR workloads. \solution decouples RL algorithm control from backend-specific execution and provides serviceized training and inference as shared cluster resources. RL jobs no longer manage private model deployments directly; instead, they issue rollout and training requests to a centralized execution substrate that manages model placement, state transitions, and scheduling across jobs. This design addresses both sides of the problem. It empowers algorithm development by allowing researchers to write against stable execution interfaces rather than framework-specific internals, and it improves cluster efficiency by turning per-job idle gaps into globally schedulable opportunities. Through affinity-aware placement, active packing, and state-aware scheduling that respect large model footprints and expensive context switching, \solution multiplexes LLM execution across jobs without frequent model migration.

We implement \solution and evaluate it on representative RLVR workloads. Our results show that \solution substantially improves effective cluster capacity while preserving the flexibility needed for rapidly evolving RL algorithms. In particular, \solution reduces user GPU-hour cost by up to 37.58\% with minimal per-job overhead, demonstrating that cluster-level multiplexing can serve as a practical systems foundation for large-scale reasoning and agentic LLM training.

Our contributions are summarized as follows:
\begin{itemize}
    \item We show that current RLVR frameworks for LLMs leave significant performance on the table, and that this limitation persists across representative \textbf{job-local execution designs}.  
    \item We present \solution, a system for \textbf{cluster-level multiplexing} of RLVR workloads that separates algorithm control from execution and enables cross-job multiplexing of rollout and training. 
    \item We design \textbf{orchestration and state-management mechanisms} that make large-model multiplexing practical in a shared cluster setting.  
    \item We implement and evaluate \solution, showing improved effective cluster capacity and reduced GPU-hour cost for RLVR workloads. 
\end{itemize}

\section{Motivation}
\label{sec:motivation}

Reinforcement learning with verifiable rewards (RLVR) has become the dominant backbone for improving LLM reasoning and agentic capabilities. Yet, despite steady algorithmic progress, practitioners still report low MFU and substantial GPU waste from idle or underloaded devices. We argue that these inefficiencies are not merely artifacts of immature implementations, but arise from a deeper structural limitation: existing systems manage resources from a user-isolated, job-local perspective.

\begin{figure}
    \centering
    \includegraphics[width=1\linewidth]{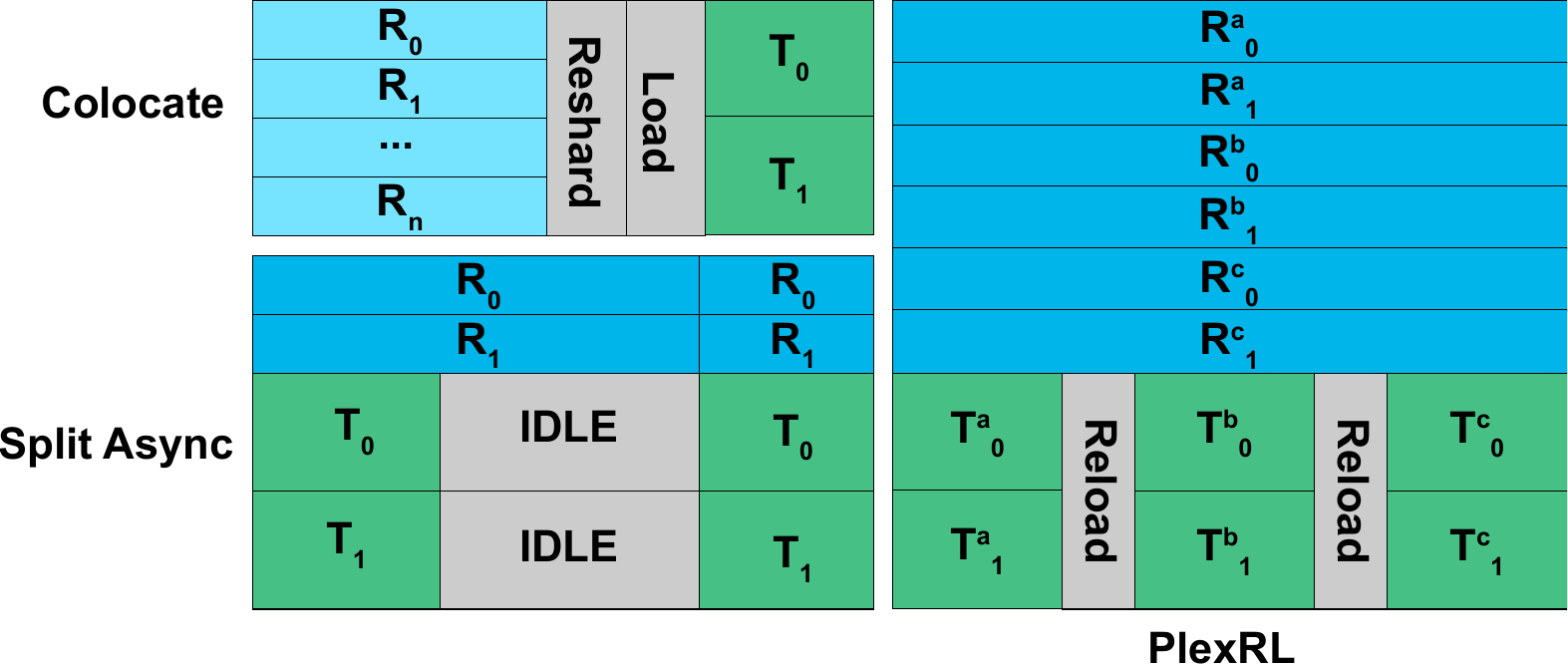}
    \caption{Common deployment patterns in RLVR Training.}
      \vspace{-1\baselineskip}
    \label{fig:pattern}
\end{figure}

We analyze three prevalent designs (Fig. \ref{fig:pattern})—split, colocated, and asynchronous split deployment—and show that all of them incur inefficiencies intrinsic to isolated job-level resource management, even though the sources of inefficiency differ across designs. These inefficiencies cannot be fully eliminated through scaling or engineering optimization within a single job. In contrast, from a cluster-level perspective, idle capacity can be recovered by allowing different jobs to complement one another’s demand gaps.

\subsection{Split Deployment: Reasonable Local MFU, Large Stall Period}

An intuitive deployment strategy is \emph{split deployment}: rollout, actor, critic, and reference models are placed on separate device pools. Each component can be deployed independently, sized according to its own workload characteristics, and orchestrated through a simple job-local script, making it convenient for researchers to prototype new algorithms, model roles, and data pipelines.

However, split deployment is highly inefficient at the job level. In the standard synchronous regime, rollout and training proceed in strict alternation: during rollout, the training pool remains idle; during actor or critic updates, the rollout pool waits; and the same pattern applies to other model roles. At best, some inference-side components such as the reference or critic model can be overlapped with rollout, but this only recovers a small fraction of the reserved capacity. The fundamental problem remains unchanged: split deployment reserves multiple disjoint device pools for a single job, while only a subset of them performs useful work at any given time.


As a result, split deployment may achieve reasonable utilization within an active phase, yet still exhibit poor utilization across the job as a whole. The inefficiency is structural rather than incidental: whenever execution is organized as phase-by-phase alternation over separately reserved pools, substantial device idling is inevitable. This makes split deployment a convenient programming model, but an extremely wasteful execution model for large-scale RLVR.

\subsection{Colocated Deployments: Large Data Parallelism, Poor MFU}

A common design is to \emph{colocate} rollout and training on the same group of accelerators. However, this computational efficiency often comes at the cost of development agility. In practice, many colocated frameworks build in strong structural assumptions: they assume a fixed set of model roles such as actor, critic, and reference; they follow a largely predetermined rollout--inference--training order; and they tightly couple algorithm logic with the underlying training and inference runtime. These assumptions enable aggressive job-local optimization, but also make the framework difficult to extend. Even minor changes to the training loop may require users to understand exactly how models are deployed, partitioned, and coordinated by the backend.

\begin{figure}
    \centering
    \includegraphics[width=1\linewidth]{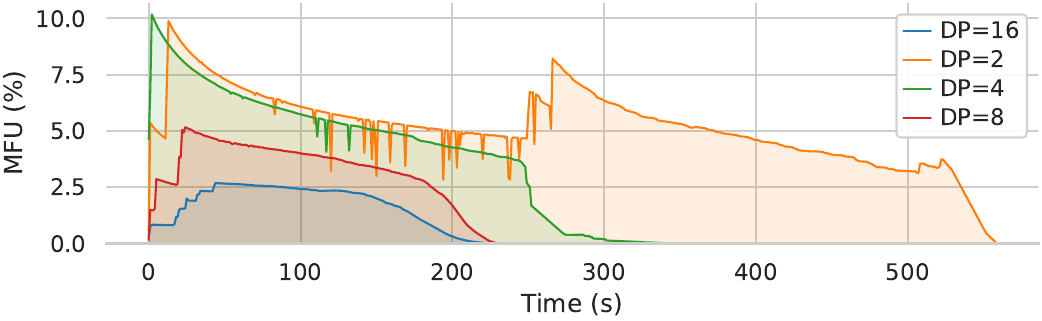}
    \caption{MFU of non-agent task under different DP size.}
      \vspace{-1\baselineskip}
    \label{fig:roll-mfu}
\end{figure}

Reusing the same devices also does not guarantee high efficiency. Training large models requires substantial per-rank state, including parameters, optimizer moments, gradients, and activation checkpoints. At hundreds of billions of parameters, a single training deployment may already span dozens to hundreds of GPUs. Rollout for the same model typically has a smaller per-replica footprint and therefore admits a lower degree of model parallelism. Under colocated execution, however, the total GPU footprint is fixed by training, so the rollout phase expands into a much larger data-parallel (DP) group than decoding can efficiently sustain. The core issue is the long-tailed nature of rollout: large DP is beneficial only during the high-batch portion of the step, but near the end, as only a few samples remain active, the effective batch size collapses and additional replicas contribute little useful work. In agentic pipelines, long-running tool calls make this tail even more pronounced by holding the rollout phase open after most requests have already finished. Consequently, many GPUs in the colocated group remain reserved through an extended low-utilization tail. (Fig. \ref{fig:roll-mfu})

In other words, colocation trades away rollout-phase MFU for the convenience of shared state and job-local execution. The larger the training footprint, the more severe this effect becomes: every long-tail sample and every delayed tool call can keep an entire oversized rollout engine reserved while doing little useful work.

\subsection{Asynchronous Rollout: Limited Slack and Inevitable Imbalance}

Given the whole-pool idleness of split deployment and the rollout inefficiency of colocated execution, asynchronous split deployment is a natural compromise. Instead of forcing rollout and training to proceed in lockstep, training continues on slightly older trajectories while rollout generates new ones in parallel. From an algorithmic perspective, this is not inherently incompatible with modern RLVR objectives: methods such as PPO, GRPO, and Reinforce++ include importance-sampling and KL-based corrections that can tolerate modest policy lag~\cite{schulman2017proximal,schulman2018highdimensionalcontinuouscontrolusing,shao2024deepseekmathpushinglimitsmathematical,hu2025reinforcestabilizingcriticfreepolicy}.

However, asynchronous execution only helps if rollout time and training time can be kept roughly balanced. If the two phases overlapped perfectly, neither side would wait for the other and device idleness would largely disappear. In practice, such balance is unattainable. There is no static device split that equalizes rollout and training time across steps: rollout suffers from sublinear DP scaling and long-tailed stalls from decoding and tool use, while training is constrained by model-parallel topology and optimizer cost. Moreover, the relative cost of the two phases varies with prompt complexity, sequence length, and tool usage. As a result, one phase remains longer than the other for most steps, so asynchronous execution can reduce waiting but cannot eliminate it.

Pushing asynchrony further also comes with an algorithmic cost. Prior work on distributed RL and RLHF suggests that excessive policy lag or heavily stale trajectories can degrade stability and sample efficiency unless they are carefully corrected and tuned~\cite{zheng2025prosperity,humayoo2025relative,munos2016safe,fujimoto2019off}. In conventional job-local frameworks, asynchronous rollout is therefore often a necessary workaround for phase mismatch, forcing developers to trade off utilization against staleness. In \solution, by contrast, asynchrony is optional: the main efficiency gain comes from reclaiming idle capacity across jobs at the cluster level, so developers can choose a more conservative synchronization regime when algorithmic sensitivity to staleness matters.

As a result, split deployment wastes devices through phase-by-phase alternation, colocated deployment wastes devices by oversizing rollout, and asynchronous rollout can only partially hide the mismatch between rollout and training time while introducing a staleness trade-off. All three designs remain confined to a job-local execution envelope: resources reserved for one job cannot be reused by another during its idle periods. This job-local binding of large model replicas to fixed resources is the root cause of cluster underutilization in RLVR workloads.

\subsection{Implication: Idle Gaps Are Unavoidable Per Job but Reclaimable at the Cluster Level}

Taken together, these observations suggest that idle gaps in RLVR are unavoidable at the level of an individual job. Split execution wastes capacity through phase alternation, colocated execution wastes capacity through oversized rollout groups, and asynchronous execution can only partially mask imbalance while introducing staleness. Since these idle periods are typically misaligned across jobs, the right optimization target is the cluster rather than the single job. Exploiting this opportunity, however, requires the system to observe and schedule fine-grained model operations rather than coarse job-level reservations. Existing RL frameworks hide these operations inside job-private deployments, coupling algorithm control to a fixed execution topology. Therefore, reclaiming idle gaps at cluster scope requires decoupling RL logic from model execution and exposing the latter through a shared execution substrate. 


\section{Insights}
\label{sec:insights}

Section 2 shows that RLVR inefficiency is fundamentally a \textbf{cluster-level reclamation problem}: idle capacity arises within jobs, but can only be recovered across jobs. This observation implies two design requirements. First, model execution must be lifted out of job-private deployments into a shared substrate visible to the cluster scheduler. Second, this substrate must preserve state locality and execution ordering so that multiplexing remains practical for large LLMs. \solution is built around these requirements. 

\noindent\textbf{Decoupling algorithm control from model execution.} Existing RL frameworks typically intertwine algorithm logic with model execution, forcing users to express rollout, evaluation, and training in terms of backend-specific operations and job-private deployments. This coupling is problematic for two reasons. First, it hides fine-grained execution opportunities from the cluster scheduler behind coarse job reservations. Second, it exposes users to intrusive integration work across specialized runtimes. In practice, the highest-performance implementations of rollout and training already come from mature systems, while RL algorithms interact with them through a comparatively stable set of primitives: generation, forward/backward computation, optimizer step, and weight synchronization. The difficulty is not invoking these primitives, but managing state exchange across runtimes efficiently. In existing frameworks, this burden often falls on users through backend-specific glue code for checkpoint conversion, synchronization, and state movement. \solution instead makes model execution a remote service with a narrow interface, so algorithm developers write only the RL logic while backend integration is absorbed by the system. 

\noindent\textbf{Cluster-Level scheduling with centralized state management.} Decoupling execution into a shared service is necessary but not sufficient. If service instances remain statically partitioned per job, the idle regions identified in Section 2 remain trapped behind the same job boundary. The execution substrate must therefore be scheduled at cluster scope. In \solution, rollout and training requests from multiple jobs are admitted by a centralized scheduler that observes request arrival, model residency, and affinity constraints, then interleaves work on a shared pool of model replicas. Because large LLM states are expensive to move, scheduling must be locality-aware: it should preferentially reuse existing placements, amortize state transitions, and avoid changing parallel layouts in the hot path. This in turn requires centralized state management. By owning model residency, offload/prefetch, checkpoint materialization, and weight synchronization, the system both makes large-model multiplexing practical and removes intrusive backend-specific state handling from user code. Together, these mechanisms let \solution reclaim idle capacity across jobs while preserving the flexibility of rapidly evolving RLVR pipelines. The next section describes how these ideas are realized in the Scheduler, execution service, and StateManager. 



\section{System Design}
\newcommand{\Router}{Router}
\newcommand{\Scheduler}{Scheduler}
\newcommand{\NodeManager}{NodeManager}
\newcommand{\RLController}{RLController}
\newcommand{\WorkerProc}{Worker}
\newcommand{\StateMgr}{StateManager}

\begin{figure}[t]
  \centering
  \includegraphics[width=\columnwidth]{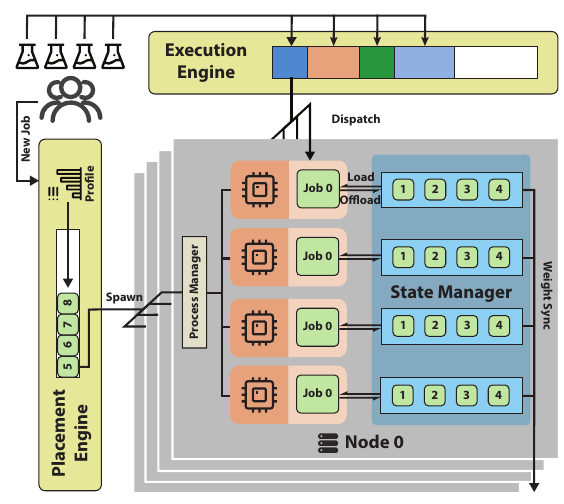}
  \caption{System design overview.}
  \label{fig:system-design}
\end{figure}

\solution transforms LLM execution for RLVR into a \emph{cluster-wide shared service} that allows large-model deployments to be safely time-multiplexed across multiple jobs. Instead of binding rollout and training to job-private model replicas, \solution exposes model execution through a remote substrate managed at cluster scope. The system is built around three requirements: (1) decoupling algorithm control from backend-specific execution, (2) multiplexing jobs while preserving affinity and per-job ordering constraints, and (3) making large-model state exchange practical through centralized management across GPU, CPU, and NVMe. To meet these requirements, \solution{} consists of three components (Figure~\ref{fig:system-design}): (i) a \textbf{\Scheduler{}} that performs cluster-wide placement and runtime ordering, (ii) a \emph{remote LLM execution service} composed of a stateless \textbf{\Router{}} and GPU-resident \textbf{\WorkerProc{}s}, and (iii) a per-node \textbf{\StateMgr{}} that manages model residency, state transitions, and checkpoint materialization. Together, these components enable multi-tenant LLM training without frequent model migration or uncoordinated memory management.

\subsection{Architecture Overview}

\RLController{} runs on CPU-only nodes and issues rollout, reward, and training requests. It holds no model state and interacts with training purely through remote calls to the execution service, allowing RL algorithms to evolve independently of the underlying LLM deployments.

\Scheduler{} performs cluster-level placement and schedules training requests from multiple jobs. It enforces each job's logical execution order while applying affinity-aware heuristics that minimize context switching and avoid unnecessary model movement. All resource arbitration happens at this layer: the execution service simply executes operations that the scheduler has admitted.

The remote LLM execution service consists of a cluster-level \Router{} and many GPU-resident \WorkerProc{} processes. The \Router{} exposes a function-oriented API, maps logical deployments to worker-process groups (WPGs), and routes operations to them. Each \WorkerProc{} is a long-lived process that owns the concrete model and optimizer and executes forward/backward/step operations on behalf of remote clients. Lightweight node-local tools such as \NodeManager{} may be used to spawn and monitor these processes but are an implementation detail rather than a core abstraction.

\label{sec:design}


\subsection{Remote Execution Abstraction}
\label{sec:remote-exec-service}

\solution{} exposes model execution through a small remote API centered on a few primitive operations: forward, forward--backward, checkpoint save/load, and weight synchronization. This interface is intentionally narrow: algorithm code specifies only the required model-side computation and state transitions, without depending on local process layout, parallelism strategy, or backend internals. As a result, RL pipelines remain decoupled from the concrete execution stack while still exposing the core operations needed for rollout-training coordination.

The basic execution unit is a \emph{worker-process group} (WPG), which represents one logical deployment of a model. A WPG consists of one \WorkerProc{} per assigned GPU and encapsulates the concrete distributed execution strategy, including data, tensor, pipeline, expert, or context parallelism. The \Router{} resolves each logical deployment identifier to its corresponding WPG and is parallelism-aware: different parallelism layouts require different input dispatch, control propagation, and result gathering patterns. Accordingly, the \Router{} is responsible for translating each admitted logical operation into the concrete communication pattern expected by the target WPG, while the backend runtime executes the actual distributed computation.

This abstraction also defines the concurrency model. For full-parameter training, operations targeting the same WPG are executed serially, yielding a single well-defined order over parameter mutation, gradient accumulation, optimizer updates, and checkpoint-visible state. Different WPGs, however, may execute concurrently when admitted by the \Scheduler{}. Thus, \solution{} enables cross-job multiplexing at cluster scope while preserving per-WPG serial semantics locally.

This narrow interface makes backend integration a systems concern rather than a user concern. Although high-performance rollout and training may rely on different specialized runtimes, their shared control surface is small: forward execution, backward/update execution, checkpoint movement, and weight synchronization. \solution{} hides backend-specific execution and state-transition mechanics behind these remote calls, allowing algorithm developers to program against a stable interface while the system handles concrete runtime integration underneath.

\begin{figure}
    \centering
    \includegraphics[width=1\linewidth]{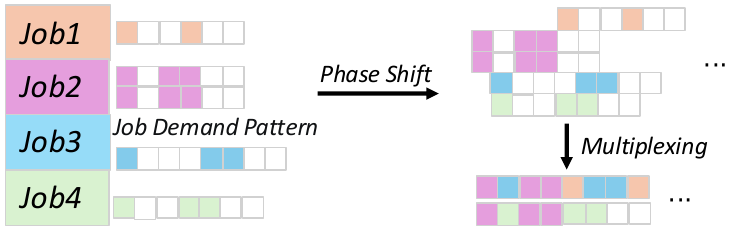}
    \vspace{-1\baselineskip}
    \caption{The job placement policy leverages job demand patterns and applies a temporal phase shift to identify the optimal node group.}
    \label{fig:NexScheduler_stage1}
\end{figure} 
\subsection{Placement}

\subsubsection{Spatio-temporal Resource Abstraction}

Effective scheduling for RLVR workloads requires reasoning over both time and placement. Because these jobs execute as recurring multi-phase cycles, the scheduler must represent future demand over a bounded horizon, quickly identify where enough cluster capacity exists, and preserve those allocations once a placement decision is made. \solution{} captures these needs through three core abstractions: 

\noindent \textbf{Cyclic Time Horizon.} The scheduler operates over a fixed-duration horizon $H$, materialized as a ring buffer over the interval $[t, t+H]$. Each job's profiled demand trace is projected onto this window, bounding the planning scope and enabling constant-space detection of contention.

\noindent \textbf{Hierarchical Resource View.} 
To tame the combinatorial search space, cluster resources are represented at two levels of granularity. 
A \textit{Global Capacity Profile} $C_{global}(t)$ tracks the aggregate number of free Nodes, enabling $O(1)$ gang-feasibility checks. Only after this macro-level pruning does the scheduler inspect \textit{Per-Node Intervals} $\mathcal{W}_p$ to perform fine-grained trace fitting. 


\noindent \textbf{Atomic Reservation.} Scheduling updates follow a commit-once semantics: once a job is placed, its footprint is immediately subtracted from $C_{global}$ across the entire cyclic horizon. This atomic reservation pre-allocates capacity for all future periods, eliminating over-commitment before the job ever begins execution.



Together, these abstractions reduce RLVR scheduling to a tractable, horizon-bounded planning problem while preserving the cyclic structure needed to reason about phase compatibility, interference, and temporal fitting across jobs.

\begin{figure}
    \centering
    \includegraphics[width=1\linewidth]{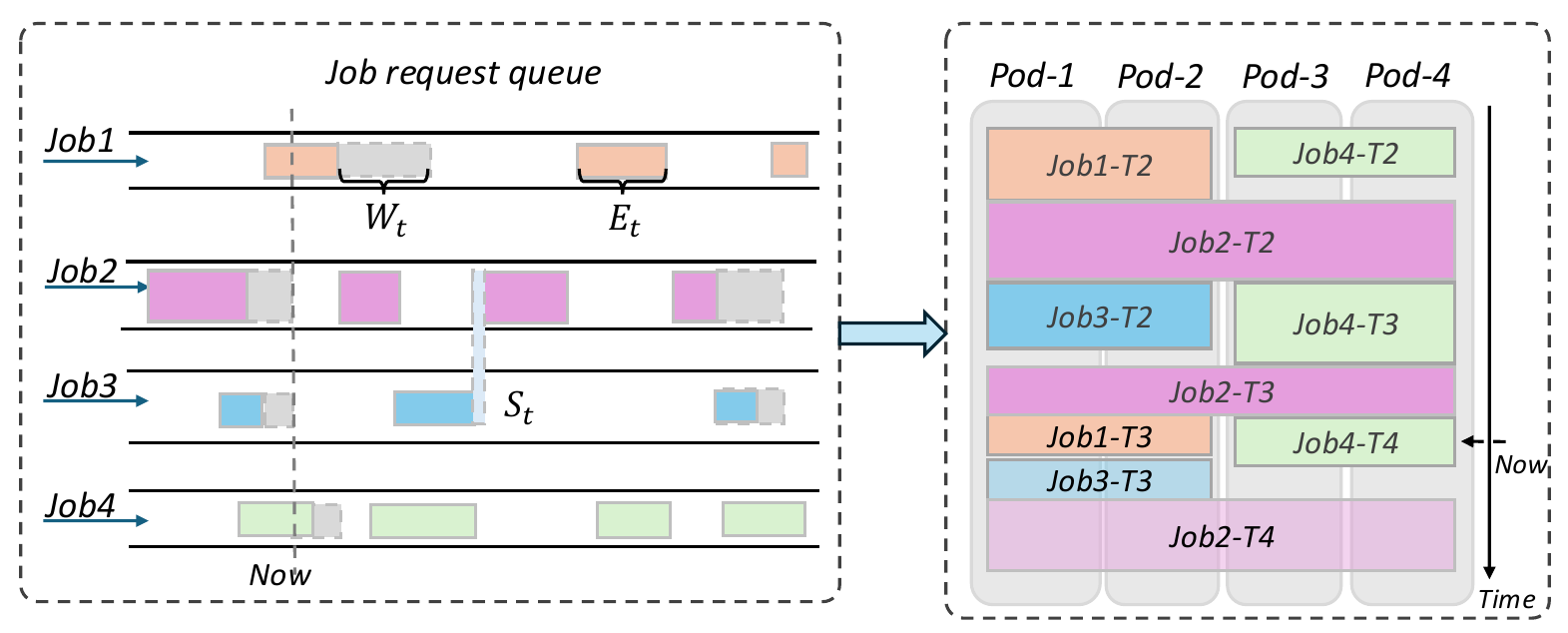}
    \vspace{-1\baselineskip}
    \caption{Request scheduling and executing.}
    \label{fig:NexScheduler_stage23}
\end{figure} 

\subsubsection{Job Placement Policy}
\label{sec:design_placement}

Given this cyclic resource view, the Scheduler chooses placements that are not only feasible but also phase-compatible. The goal is to pin a job to a node group whose existing demand pattern leaves enough slack at the right times to absorb the new job with minimal interference, while still avoiding costly model initialization and movement. To this end, \solution{} enforces stable node-group placement and adopts a dual-phase policy conditioned on trace availability.  \noindent (1) \textbf{Cold Start.} For jobs without historical traces, the Scheduler provisions dedicated Node groups using the \textit{Global Capacity Profile.} This isolation prevents cross-job interference and allows the profiler to extract clean execution signatures, enabling accurate modeling of future demand patterns.
(2) \textbf{Warm Start.} 
For jobs with valid trace profiles, the profiling phase is skipped. The Scheduler directly consults the \textit{Hierarchical Resource View} to eliminate globally infeasible windows, then performs fine-grained placement by fitting the job’s demand pattern into the \textit{Per-Node Intervals} of eligible node groups.


Once a full resource-usage cycle is profiled, the scheduler triggers a repacking event to improve packing density. As shown in Figure~\ref{fig:NexScheduler_stage1}, placement is cast as a spatio-temporal trace-fitting problem: the job's periodic demand trace is shifted within the scheduler's planning horizon and fitted against the free windows of candidate node groups.


Let $S = \{(a_i, d_i)\}_{i=1}^N$ denote the discrete \textit{execution segments} of job $J$, where $a_i$ is the relative offset and $d_i$ is the duration within period $T$. For a candidate node group, let $\mathcal{W}$ denote the set of disjoint free windows available to that group over the planning horizon.
The scheduler seeks a \textit{Micro-Shift} $\delta \in [0, \alpha T]$ that maps all segments into $\mathcal{W}$ while minimizing latency. This approach effectively identifies "micro-slots" in the cluster timeline. The objective is to minimize the \textit{Scheduling Cost}:
\vspace{-0.5\baselineskip}
\begin{equation}
\min_{\delta} \quad \mathcal{J}(\delta) = w_1 \cdot \underbrace{\frac{t_{end}(\delta) - T}{T}}_{\text{Completion Delay}} + w_2 \cdot \underbrace{\frac{\delta}{T}}_{\text{Start Shift}}
\label{eq:micro_shift_cost}
\end{equation}

\noindent subject to the constraint that each shifted segment fits entirely within a free window:
\begin{equation}
\forall (a_i, d_i) \in S, \exists [s, e) \in \mathcal{W} : s \le (a_i + \delta) \land (a_i + \delta + d_i) \le e
\end{equation}

When multiple candidate node groups satisfy this fitting constraint, the Scheduler further ranks them by predicted phase interference, favoring placements whose shifted active segments align with slack regions of resident jobs rather than with their critical phases.

\subsection{Runtime Scheduling}
Placement determines which jobs share a node group; runtime scheduling determines whether that sharing remains stable over time. A purely reactive queueing policy is insufficient for RLVR workloads because each job's requests arise from a cyclic dependency chain: the arrival time of an operation depends on the completion time of earlier operations in the same job. Naively inserting foreign work whenever capacity becomes available can therefore keep the system in a persistently perturbed regime, where jobs repeatedly delay one another's critical phases and induce unnecessary switching. \solution{} instead performs critical-path-aware runtime scheduling: it profiles each job's cycle, identifies where slack naturally exists, and grants execution preferentially at timings that fit within these slack regions.
\begin{algorithm}[h]
\caption{Request Schedule with HRRS}
\label{alg:timeline_replanning}
\small
\SetKwInOut{Input}{Input}
\SetKwInOut{Output}{Output}

\Input{New Request $R_{new}$, Current Resource View $\mathcal{V}$ (contains $R_{running}$ and List of $R_{scheduled}$)}
\Output{Updated Resource View $\mathcal{V}'$ with new timeline}

$\Omega \leftarrow \{R_{new}\} \cup \{R_{running}\} \cup \mathcal{V}.get\_scheduled\_requests()$\;

\For{$R \in \Omega$}{
    $t_{wait} \leftarrow t_{now} - R.arrival\_time$\;
    \If{$R == R_{running}$}{
        $t_{req} \leftarrow R.remaining\_time$\;
    }
    \Else{
        $t_{req} \leftarrow R.exec\_time + t_{load} + t_{offload}$\;
    }
    $R.score \leftarrow (t_{wait} + t_{req}) / t_{req}$\;
}

Sort $\Omega$ by $R.score$ \textbf{descending}\;

$t_{cursor} \leftarrow t_{now}$\;
$\mathcal{V}'.clear()$\;

\For{$R \in \Omega$}{
    \If{$R \neq R_{running}$ \textbf{and} $t_{cursor} == t_{now}$}{
        $t_{cursor} \leftarrow t_{cursor} + t_{offload} + t_{load}$\;
        Stop $R_{running}$ if exists\;
    }
    $t_{start} \leftarrow t_{cursor}$\;
    $t_{end} \leftarrow t_{start} + R.req\_time$\;
    $\mathcal{V}'.assign(J, [t_{start}, t_{end}])$\;
    $t_{cursor} \leftarrow t_{end}$\;
}

\Return $\mathcal{V}'$\;
\end{algorithm}

Ideally, the placement strategy guarantees conflict-free interleaving of jobs. In practice, strict non-overlap is infeasible due to allowable interference thresholds and inherent execution jitter. Under these conditions, a naive First-Come, First-Served (FCFS) approach exhibits two major shortcomings:

\begin{itemize}
\item \textbf{Excessive Switching Overhead.} Context switching requires migrating model states between GPU HBM and host memory. Frequent loading and offloading accumulate significant overhead, severely degrading aggregate GPU utilization.
\item \textbf{Head-of-Line Blocking.} Short tasks arriving shortly after long-running operations suffer disproportionate wait times, negatively impacting system-wide responsiveness.
\end{itemize}

To address these challenges, we propose \textit{Highest Response Ratio with Setup} (HRRS). HRRS extends the classical HRRN algorithm by explicitly incorporating the setup cost into the priority calculation as shown in Algorithm \ref{alg:timeline_replanning}.
For a pending task $i$ arriving at time $t_a$, let $W_i(t) = t - t_a$ be its wait time, and $E_i$ be its estimated execution time. We define the \textit{Effective Service Time} $S_i(t)$ as:
  \vspace{-0.5\baselineskip}
\begin{equation}
S_i(t) = E_i + \mathbf{1}_{\text{switch}}(i, \text{curr}) \cdot (T_{offload} + T_{load})
\label{eq:effective_service}
\end{equation}

where $\mathbf{1}_{\text{switch}}$ is an indicator function that equals 1 if task $i$ requires a context switch from the currently running task, and 0 otherwise. The dynamic priority $P_i(t)$ is then given by:
  \vspace{-0.5\baselineskip}
\begin{equation}
P_i(t) = \frac{W_i(t) + S_i(t)}{S_i(t)} = 1 + \frac{W_i(t)}{E_i + \mathbf{1}_{\text{switch}} \cdot \mathcal{C}{setup}}
\label{eq:hrrs_priority}
\end{equation}

By inflating the denominator when $\mathbf{1}_{\text{switch}}$, HRRS naturally encourages \textit{batching} of similar tasks to amortize setup costs, while ensuring that long-waiting tasks eventually gain high priority to prevent starvation. The remaining setup cost is further reduced by \StateMgr{}, which can execute many state-management actions off the GPU critical path; we describe these mechanisms in \S\ref{sec:nexcache}.

\subsection{Model State Manager}
\label{sec:nexcache}

Cluster-level multiplexing requires more than scheduling jobs onto shared accelerators: it also requires a mechanism that maps abstract scheduling decisions onto the concrete state layout of a node. In \solution{}, this role is served by a per-node \StateMgr{}. Unlike the \Scheduler{}, which reasons over a cluster-wide virtual resource view, or the execution service, which operates at the granularity of logical deployments, \StateMgr{} maintains a \emph{resource-resident view} of model state on each node. It is the bridge between virtual scheduling decisions and hardware-bound state scope: it knows which tensors are resident in GPU memory, which have been offloaded to host memory or NVMe, which states are shared across ranks, and what movement or materialization is required before a deployment can execute. By making this node-local physical view explicit, \solution{} can safely time-multiplex large models without relying on opaque per-process memory management.

\begin{figure}
    \centering
    \includegraphics[width=1\linewidth]{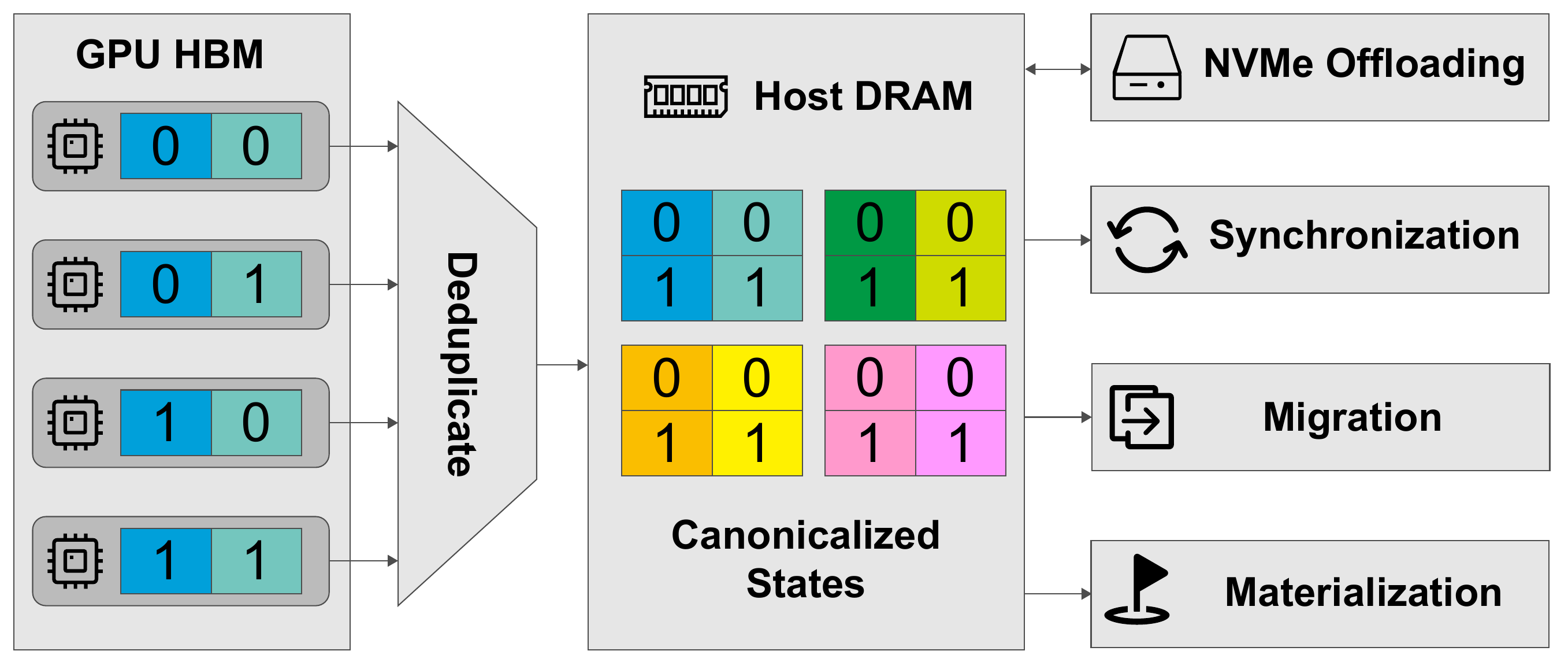}
    \vspace{-1\baselineskip}
    \caption{Schematic of the Model State Manager.}
    \label{fig:statemgr}
\end{figure}

\subsubsection{Hierarchical Residency}
\StateMgr{} centrally manages model and optimizer state across a three-tier hierarchy: GPU memory for active execution, host memory for warm offloaded state, and NVMe for cold spillover. This hierarchy is essential because cluster-level scheduling decisions are only meaningful if they can be realized within the actual capacity of each node. If each worker process were allowed to manage offloading independently, the \Scheduler{} would reason against an inaccurate abstract view while memory pressure, fragmentation, and I/O interference accumulated below it. By contrast, \StateMgr{} provides a single node-local authority over residency, eviction, and admission. This lets the \Scheduler{} treat placement and interleaving as logical decisions while delegating the concrete realization of state residency to a component that sees the true hardware-bound picture. It also enables scheduler-directed prefetching: when an upcoming context switch is predicted, \StateMgr{} can proactively move state upward in the hierarchy before the corresponding deployment becomes active.
\subsubsection{Canonicalized Offloaded State}
A second responsibility of \StateMgr{} is to provide a canonical representation of offloaded state. In large distributed training jobs, process-local memory images often contain substantial redundancy, especially under data-parallel replication where logically identical tensors appear on multiple ranks. Offloading such state naively would duplicate storage, inflate host and NVMe pressure, and make subsequent reconstruction dependent on process-local layout. \solution{} avoids this by indexing offloaded tensors by logical keys rather than by process ownership. This allows \StateMgr{} to deduplicate shared state across replicas while preserving enough metadata to reconstruct the tensor views required by a particular deployment. As a result, offloaded state is stored in a normalized node-local form rather than as a collection of opaque worker snapshots. This canonicalized representation is critical to making offloading, reuse, and migration practical across heterogeneous parallel configurations.
\subsubsection{Materialization, Synchronization, and Migration}
Because \StateMgr{} owns both the physical placement and canonical representation of state, it can also perform the transformations needed by other system components. First, it supports \emph{transparent checkpointing}: checkpoint creation is treated as materialization from managed state rather than as an explicit user-triggered export path. Even when part of the state is offloaded, \StateMgr{} can resolve the relevant tensors into checkpoint shards directly. Second, it supports weight synchronization to rollout deployments by materializing training-visible state into the format expected by serving instances. Third, it enables cross-node deployment migration by mirroring managed state to a destination node and reconstructing the target deployment there. In all three cases, the key principle is the same: state transformations are owned by the system layer rather than reimplemented in user code or backend-specific control paths.

\subsubsection{Overlapping State Management with Execution}
Centralizing node-local states allows many state-management operations to proceed off the GPU critical path. Once a stable offloaded copy exists, \StateMgr{} can operate on managed host-side state while unrelated WPGs continue executing; for example, the optimizer step can run on host-resident state when CPU optimizer~\cite{ren2021zerooffloaddemocratizingbillionscalemodel} is enabled, checkpoint shards can be materialized from managed offloaded state without interrupting training, and state can be prefetched or drained across the memory hierarchy asynchronously. As a result, only operations that directly access or mutate the active GPU-resident deployment remain on the critical path, helping \solution{} multiplex large models without turning each context switch into a full stall.

Taken together, these mechanisms make \StateMgr{} the node-level state authority for \solution{}. The \Scheduler{} decides \emph{which} deployment should run and \emph{when}; \StateMgr{} determines whether the required state is resident, how it should be materialized, and what movement must occur to realize that decision on hardware. This separation is what makes large-model multiplexing practical without frequent full redeployment or uncoordinated memory management.

\section{Implementation}
\subsection{Remote Execution Service}
\label{sec:remote-exec-impl}
We implement the remote execution service as a thin control/data-plane split. A stateless Router serves as the control-plane entry point for remote model operations issued by \RLController{}. Rather than dispatching requests directly to execution backends, the Router first submits them to the Scheduler, which decides when the target deployment may run; only admitted operations are then forwarded to the corresponding WPG. This design preserves the function-oriented interface described in Section \ref{sec:design} while ensuring that all execution remains subject to cluster-level ordering and affinity decisions.

On the data plane, each WPG consists of one \WorkerProc{} per assigned GPU and is backed by an existing distributed training runtime. We implement \WorkerProc{} as thin adapters around Megatron- or FSDP-based ranks: a Worker receives an admitted operation from the Router, translates it into the corresponding backend call, executes it within the local runtime, and returns results to the caller. \solution{}therefore does not replace the backend's distributed execution mechanisms. Data, tensor, pipeline, expert, and context parallelism remain entirely managed by the underlying runtime, while the remote execution layer controls only admission, dispatch, and operation ordering.

The Router maintains a logical mapping from deployment identifiers to WPGs and handles deployment lifecycle events such as worker-group creation, model initialization, and checkpoint loading. These mechanisms are intentionally lightweight: they provide a stable namespace and invocation path for remote execution without exposing backend-specific process layout to the RL algorithm. As a result, algorithm code interacts with a narrow remote interface, while execution details remain encapsulated inside the service layer.

Although the Router accepts requests from many clients concurrently, execution is serialized within each WPG in the full-parameter training setting. This yields a well-defined order over parameter mutation, gradient accumulation, optimizer updates, and checkpoint-visible state, preventing interference between overlapping requests to the same deployment. Different WPGs, however, may execute concurrently when admitted by the Scheduler. This combination matches the design goal of \solution: cluster-wide multiplexing across deployments, while preserving local execution semantics within each deployment.

\subsection{Scheduler}
\subsubsection{Spatio-Temporal State Management}
 To avoid the $O(N \cdot T)$ cost in Eq. \ref{eq:micro_shift_cost}, we utilize a hierarchical indexing structure combined with a signal-based profiling pipeline.

\noindent \textbf{Circular Ring Buffer.} 
 The timeline is mapped to a fixed-size \textit{Ring Buffer} ($\mathcal{T}$, 28,800 slots for a 8-hour horizon). Modulo arithmetic ($t_{idx} = t_{abs} \pmod{L}$) supports an unbounded horizon without shifting the array.

\noindent \textbf{Segment Tree Pruning.} 
We maintain a global segment tree over the ring buffer to support $O(\log T)$ Range Minimum Queries. 
For a job requiring $K$ Nodes, the Scheduler verifies $\min_{t \in [t_{now}, t_{now}+d]} (\text{Capacity}(t)) \ge K$ before checking specific Nodes. 
This instantly prunes infeasible time windows—filtering out over 80\% of the search space—before accessing granular states.

\noindent \textbf{Interval Set Fitting.} 
For candidate Nodes, we track resource availability using Interval Sets (sorted disjoint free ranges $[s_i, e_i)$) to compress memory footprint. 
Trace fitting is implemented as a binary search (\texttt{bisect}) over interval boundaries (\texttt{simulate\_insert}). 
This efficiently verifies if time-shifted segments fit into free windows in $O(\log M)$ time, avoiding expensive linear scans.

\subsubsection{Asynchronous Schedule Plan}

We implemented a non-blocking control plane to manage high-concurrency requests and context switching.

\noindent \textbf{Non-blocking Request Handling.} To handle API requests without blocking, the \texttt{submit\_queued\_operation} routine wraps each request in a \texttt{QueuedOperation} object containing an \texttt{asyncio.Future} handle. These objects are immediately pushed into per-job \texttt{request\_queues}. This allows the API handler to return immediately while the scheduler processes the operations in the background.

\noindent \textbf{Automatic Context Switching.} To manage model swapping, the scheduler maintains a map (\texttt{group\_executor\_gpu\_job}) tracking the Job ID currently resident on each GPU group. In the \texttt{\_handle\_job\_transition} function, the system compares the incoming operation's target Job ID with this map. If they differ, the system automatically prepends offload and load operations to the execution stream, ensuring the correct model is loaded before the user command runs.

\subsubsection{Task Executor}


The Task Executor is implemented as the operational backbone of the Scheduler, translating logical placement decisions into concrete lifecycle management. To handle asynchronous execution and cluster-level coordination, the Executor maintains a lightweight finite state machine (FSM) for each job, managing state transitions through the following mechanics:

\noindent \textbf{Priority-based Admission} (\texttt{QUEUED}). Upon submission, jobs are injected into a pending pool. Rather than simple FIFO processing, the Executor continuously evaluates queued jobs against dynamic cluster resource availability, actively computing their HRRS scores to dictate the admission order.

\noindent \textbf{Lock-Gated Execution} (\texttt{RUNNING}). The transition to active execution is strictly concurrency-controlled. To prevent resource collisions during model swapping, the Executor implements a gating mechanism: a job transitions to the running state only after prerequisite tasks finish and the Executor successfully acquires the exclusive lock for the designated training services Node. 

\noindent \textbf{Lifecycle Teardown} (\texttt{COMPLETED}). Once the job finishes execution, the FSM finalizes the lifecycle, triggering the safe release of node locks and signaling the Scheduler to unblock subsequent operations in the pipeline.

\subsection{Implementation: Model State Manager}

We implement \StateMgr{} as a per-node sidecar daemon that mediates all state movement between GPU memory, pinned host memory, and NVMe. This sidecar design preserves the centralized residency model from Section \ref{sec:design} without entangling memory management with backend-specific worker logic. Workers interact with \StateMgr{} through a thin client interface that exposes blocking state-transfer primitives. We intentionally keep this interface synchronous: when a transfer returns, the requested state is resident and safe to use on the default CUDA stream. This avoids pushing stream-synchronization and event-management complexity into the training or rollout codepaths.

On the server side, \StateMgr{} separates a lightweight control plane from a data plane specialized for tensor movement. Metadata and coordination requests are handled asynchronously, while data transfers use a simple synchronous protocol over Unix domain sockets together with CUDA IPC handles. In our design, the sidecar acts as the active transfer endpoint: it maps the client-exported device memory, performs the requested copy, and signals completion only after the transfer has finished. This centralizes CUDA context management inside the daemon and avoids per-client coordination hazards such as premature handle destruction or inconsistent stream ordering. To support hierarchical residency efficiently, the host tier uses pinned memory, while the NVMe tier bypasses the page cache through direct I/O.

In conventional split deployments, weight exchange between training and rollout commonly goes through checkpoint files written to and read from shared storage. We instead build checkpointing and weight synchronization directly on top of \StateMgr{}. Because \StateMgr{} already owns the canonical offloaded state, it can materialize rollout-visible shards directly from managed memory and serve them over RDMA-capable interconnects without first flushing them to disk. ... Crucially, resharding is performed on the fly and without redundant transfer: each rollout rank fetches only the tensor slices required by its target parallel layout, rather than pulling a full tensor or checkpoint replica. This zero-redundancy design is essential at large model scales, where reconstructing full tensors would both inflate transfer volume and risk OOM during synchronization.

\begin{figure*}[t]
  \centering
  \includegraphics[width=\textwidth]{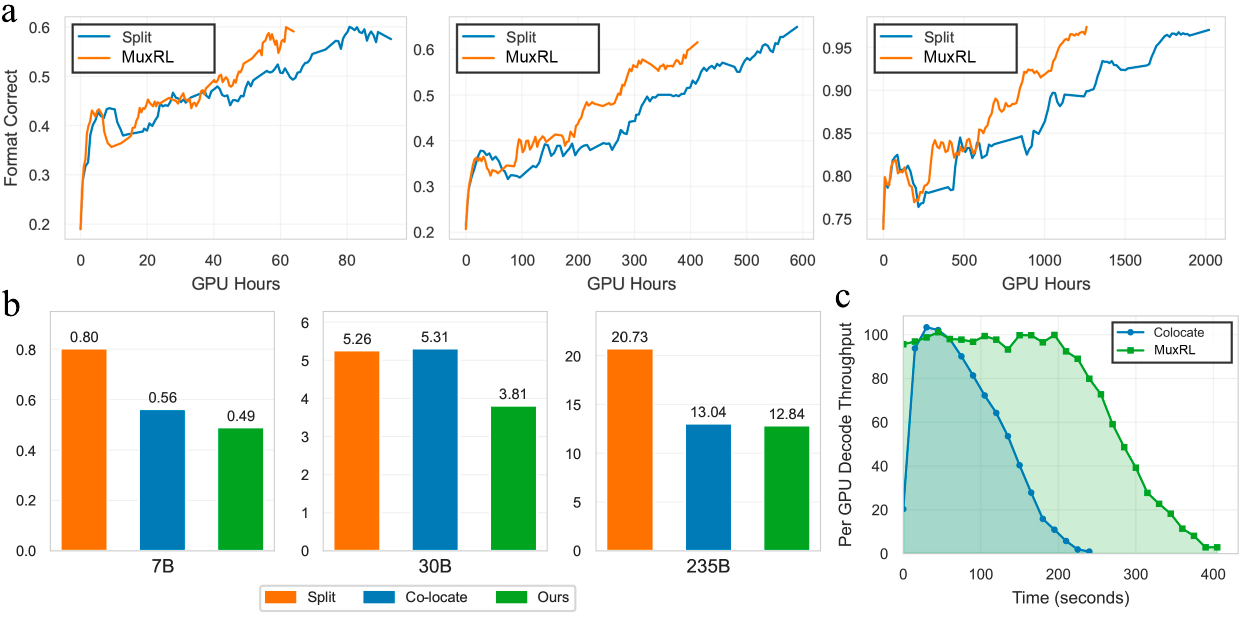}
  \caption{End-to-end evaluation of \solution in mathematical task. \textbf{a,} Reward dynamics over training. \textbf{From left to right:} 7B, 30B, 235B; \textbf{b,} GPU hour cost per step; \textbf{c,} Decoding throughput per GPU under colocated (large DP) and \solution (small DP) settings. Snapshot taken from same steps of 235B model training.}
  \label{fig:e2e-conv}
\end{figure*}
\section{Evaluation}
\subsection{Setups}
\label{sec:eval-setup}
Our evaluation is conducted on a 2048-GPU cluster with Kubernetes resource management. To validate \solution in a real RLVR scenario, we performed end-to-end training on a proprietary mathematical problem dataset with 5 difficulties comparable to AIME, consisting of around 45,000 samples. To examine effectiveness across different scales and architectures, we employed Qwen2.5-7B-Instruct (Dense), Qwen3-30B-A3B-Thinking-2507 (Mixture-of-Experts), and Qwen3-235B-A22B-Instruct-2507 (Mixture-of-Experts). The model parallel settings refer to Tab. \ref{tab:train_par}. DP, CP, EP, PP, TP refer to data, context, expert, pipeline, and tensor parallel size, respectively. Subscript $R$ denotes rollout configuration. ZeRO stage 2 are applied for all trials. Additionally, ZeRO-offload is enabled for 235B model to effectively accommodate the model states.

\subsection{End-to-End Cost Efficiency}
\label{sec:eval:e2e}

We begin by measuring the end-to-end effect of \solution on RLVR training efficiency. As detailed in \S\ref{sec:eval-setup}, all experiments use the same math-oriented RLVR workload and identical algorithms and hyperparameters across system configurations. For each model scale, we evaluate three deployment regimes: (i) a colocated baseline, (ii) a split asynchronous (Split-Async) baseline, and (iii) \solution, where rollout uses per-job GPUs while training GPUs form a shared pool that is time-sliced across two concurrently active jobs. GPU allocations for all settings are listed in Tab.~\ref{tab:train_par}.

Our primary efficiency metric is \emph{GPU-hours per effective training step}, computed as the total GPU time (training + rollout) divided by the number of completed training steps. Fig.~\ref{fig:e2e-conv}a reports model reward progress against consumed GPU-hours. For a fair comparison, \solution is evaluated under two-job packing; the colocated baseline is excluded from the asynchronous trace plots because it cannot support async rollout. As expected, \solution preserves training quality---reward trajectories match those of the baselines, consistent with the fact that \solution does not alter algorithmic semantics. Under identical 100-step budgets, two-job packing reduces end-to-end GPU-hours by \textbf{31.36\%, 30.10\%, and 37.58\%} compared to asynchronous split deployment for the three model sizes.

A more nuanced trend appears when examining GPU-hour efficiency at matched response lengths (Fig.\ref{fig:e2e-conv}b). We observe increasing end-to-end cost for the colocated regime as model size grows. For the 30B model, even bubble-free colocated execution becomes more expensive than Split-Async. This behavior arises from the escalating \emph{context-switch cost} between rollout and training modes: loading optimizer states from host RAM to GPU takes \textbf{19.0 seconds}, and this overhead grows with model size. The pattern is further validated in the 235B experiment: enabling ZeRO-Offload---placing optimizer states in host memory---drastically reduces colocate overhead relative to Split-Async, since device--host transfers are eliminated; however, this comes at the cost of a longer optimizer step.

Even with only two jobs packed, \solution retains a substantial advantage over colocated execution because colocated rollout forces large DP sizes that cannot saturate GPU throughput. Fig.~\ref{fig:e2e-conv}c highlights this effect: in small-DP settings, \solution achieves a \emph{real-throughput AUC / peak-throughput AUC} ratio of \textbf{75.03\%}, compared to \textbf{52.74\%} under colocated deployment. This disparity directly reflects the \emph{unsaturated computation} induced by the oversized rollout DP required by colocated designs, leading to significant GPU underutilization.

\begin{table}[h]
\caption{Parallel Settings.}
\centering
\label{tab:train_par}
\begin{tabular}{@{}lllllll@{}}
\toprule
\textbf{Model Size} & \textbf{$DP$} & \textbf{$CP$} & \textbf{$EP$} & \textbf{$PP$} & \textbf{$DP_R$} & \textbf{$TP_R$} \\ \midrule
\textbf{7B}         & 2           & 4           & NA          & 1           & 1              & 2              \\
\textbf{30B}        & 8           & 8           & 8           & 1           & 4              & 2              \\
\textbf{235B}       & 1           & 8           & 8           & 12          & 4              & 8              \\ \bottomrule
\end{tabular}
\end{table}

\subsection{Unleashing cluster capacity via spatio-temporal packing.}

We collected RL job statistics from three months of active cluster operation and replayed them in a trace-driven simulation to estimate how much capacity \solution{}can recover. To preserve realism, simulated requests follow a typical agentic GRPO setup: (i) requests are mapped to similarly sized RL tasks using execution characteristics extracted from Weights \& Biases traces; (ii) each function invocation within a job executes strictly serially; and (iii) asynchronous rollout permits one step of staleness, with synchronization enforced at the end of each iteration.

We compare four scheduling policies: \textit{Isolated}, \textit{Pack}, \textit{Spread}, and \textit{Spread+Backfill}. Figure~8 shows the cumulative distribution of normalized queueing delay, measured as \texttt{wait\_time} / \texttt{job\_duration}, together with the total time required to complete the workload. The \textit{Isolated} baseline has a much heavier tail than all other policies: many jobs wait several multiples of their own execution time, with worst cases far beyond those under shared scheduling. In contrast, all three sharing-aware policies substantially compress the delay distribution, showing that much idle capacity can be reclaimed once scheduling is lifted from the job level to the cluster level.

Among the shared policies, \textit{Pack} removes most pathological queueing, while \textit{Spread} further improves robustness by reducing contention between simultaneously active phases. \textit{Spread+Backfill} performs best overall, especially in the low- and medium-delay regime, indicating that backfilling effectively reclaims short residual slack windows left after coarse-grained placement. Put differently, phase-aware spreading captures most large idle gaps, and backfilling converts the remaining fine-grained fragmentation into useful work.

This gain is also visible at the workload level. As shown in Figure~8, \textit{Spread+Backfill} reduces total completion time to 56.0\% of the \textit{Isolated} baseline, a 44.0\% reduction in time-to-drain the trace. Equivalently, the same cluster can sustain roughly 1.8$\times$ more RLVR workload under the same capacity budget. These results support the central claim of \solution: the main inefficiency in RLVR clusters is not a shortage of raw accelerators, but fragmentation from job-local reservation and phase misalignment. Once placement becomes spatio-temporal and residual holes are backfilled, stranded capacity is converted directly into useful training progress.

\begin{table}[h]
    \centering
    \caption{Bubble Ratio Analysis across Models}
    \label{tab:bubble_ratio}
    \begin{tabular}{lrrr}
        \toprule
        \textbf{Metric} & \textbf{7B} & \textbf{30B} & \textbf{235B} \\
        \midrule
        Cycle Time (s)          & 289.03 & 284.80 & 589.71 \\
        \midrule
        compute\_log\_prob (s) & 9.66  & 19.62  & 20.11 \\
        update\_actor (s)      & 38.08  & 56.35  & 82.39 \\
        sync\_weight (s)   & 9.76   & 7.57   & 8.89 \\
        \midrule
        \textbf{Bubble Ratio}   & \textbf{80.10\%} & \textbf{70.67\%} & \textbf{81.11\%} \\
        \bottomrule
    \end{tabular}
\end{table}

\section{Discussion}
\subsection{Emergent Relaxation of \solution}
As discussed in Section~\ref{sec:motivation}, asynchronous rollout introduces a mismatch between training and rollout durations, which often leaves slack in the device utilization timeline: accelerators are reserved for a job but remain idle for short intervals. This slack can safely absorb additional function execution latency without affecting the job’s completion time. However, because RL pipelines are largely serial---with few exceptions in multi-model settings such as PPO-style multi-policy training or distillation---any injected delay in one phase eventually propagates to subsequent phases. Once the cumulative delay within a step exceeds the available slack, the overrun spills into later steps, turning former idle intervals into active ones and ultimately increasing the end-to-end completion time.

Our key observation is that this chained delay creates an opportunity rather than merely a liability. When multiple jobs are colocated, the propagated stalls naturally phase-shift their rollouts relative to one another. Our placement policy exploits this effect by preferentially bundling jobs with similar or commensurate step periods. As these jobs interact, the induced delays gradually shift their timelines into a low-interference equilibrium: the idle gaps of one job are increasingly aligned with the active phases of another. As a result, each additional unit of task delay contributes less and less to actual overlap (and thus slowdown) for any single job, effectively amortizing per-task interference and improving aggregate device utilization without explicit fine-grained coordination.

\subsection{Service Level Objective of LLM Training Service}

Systems supporting LLM inference have well-established service level objectives (SLOs), such as time-to-first-token and steady-state decoding latency, but analogous SLOs for LLM training remain largely undefined. This absence poses an immediate challenge for systems like \solution. Although aggressively packing jobs maximizes cluster-level MFU, it can also inflate job completion time far beyond reasonable bounds, directly contradicting our goal of enabling agile algorithm development.

To prevent pathological oversubscription, we introduce an explicit upper bound on acceptable slowdown. In this work, we use the \emph{duty ratio} of the training timeline as a practical indicator: once the fraction of time a job spends waiting exceeds this bound, additional colocated jobs are rejected. From the user’s perspective, the primary objective is still to minimize the end-to-end cost of RLVR training. \solution achieves this by substantially reducing GPU-hours per task. Unified provisioning of training deployments ensures that users are billed strictly for the computation they consume, rather than overpaying for monolithic, per-job training stacks.


\begin{figure}
    \centering
    \includegraphics[width=0.9\linewidth]{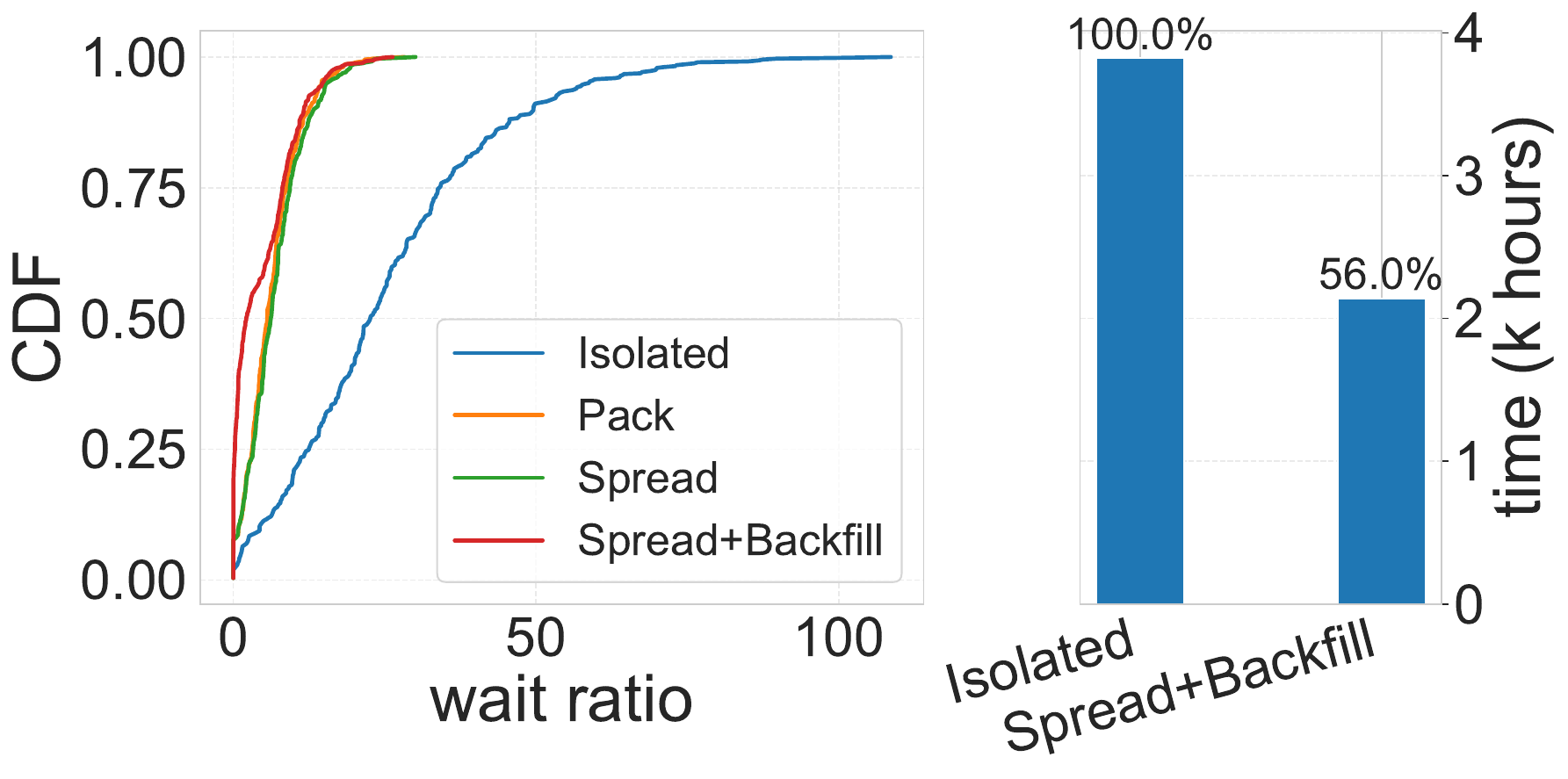}
    \vspace{-1\baselineskip}
    \caption{CDF of job queueing delay and makespan comparison across scheduling policies.}
    \label{fig:wait_task}
    \vspace{-1\baselineskip}
\end{figure}

\section{Related Work}

Prior systems improve different parts of the LLM execution stack. On the inference side, systems such as FastServe~\cite{wu2024fastdistributedinferenceserving}, Llumnix~\cite{sun2024llumnixdynamicschedulinglarge}, vLLM~\cite{kwon2023efficientmemorymanagementlarge}, and ServerlessLLM~\cite{ServerlessLLM} study scheduling, memory management, request migration, and cold-start reduction for LLM serving. At the cluster level, systems such as Gandiva~\cite{gandiva}, Tiresias~\cite{tiresias_gu2019}, Gavel~\cite{gavel_narayanan2020}, Salus~\cite{salus_yu2020}, and PipeSwitch~\cite{pipeswitch_bai2020} show the benefits of GPU sharing, time-slicing, and fast switching for deep learning workloads.

For alignment workloads, frameworks such as TRL~\cite{vonwerra2022trl}, OpenRLHF~\cite{hu2025openrlhfeasytousescalablehighperformance}, DeepSpeed-Chat~\cite{yao2023deepspeedchateasyfastaffordable}, NeMo-Aligner~\cite{shen2024nemoalignerscalabletoolkitefficient}, and veRL~\cite{sheng2024hybridflow} improve programmability and throughput for RLHF/RLVR pipelines. Recent systems such as ReaLHF~\cite{realhf_mei2024} and RLHFuse~\cite{rlhfuse_zhong2025} further optimize execution within a single RLHF job through better dataflow planning and finer-grained coordination across stages.

\solution addresses a different systems boundary. Existing serving and scheduling systems do not target multi-stage RLVR training, while existing RLHF/RLVR frameworks largely optimize within a job-local execution boundary. In contrast, \solution decouples RL control from model execution and exposes rollout and training as a shared cluster-managed substrate. This enables multiplexing across RLVR jobs and targets a source of inefficiency that job-local optimizations alone cannot remove.



\section{Conclusion}

In this paper, we argued that the inefficiency of RLVR training is fundamentally a multi-tenant problem rather than only a job-local optimization problem. Individual RLVR jobs inevitably contain idle periods caused by phase alternation, long-tailed rollout behavior, tool stalls, and asymmetric rollout--training resource demands. These gaps are difficult to eliminate within a single job, but they are often misaligned across jobs and therefore recoverable at cluster scope. \solution exploits this opportunity by decoupling RL algorithm control from model execution and exposing training and inference as shared, schedulable services. Its scheduler and StateManager make this practical for large models by preserving affinity, reducing disruptive switching, and managing model state across GPU memory, host memory, and NVMe.

Our evaluation shows that this design yields substantial practical benefit. Across representative RLVR workloads, \solution reduces end-to-end GPU-hour cost by up to 37.58\% while preserving training behavior. More broadly, these results suggest that future RL systems should treat cluster-visible execution and centralized state management as first-class design concerns, rather than relying solely on increasingly specialized job-local runtimes. As RLVR and agentic training continue to diversify, cluster-level multiplexing offers a compelling systems foundation for combining efficiency with algorithmic flexibility.

\bibliographystyle{ACM-Reference-Format}
\bibliography{reference}

@misc{deepseekai2025deepseekr1incentivizingreasoningcapability,
      title={DeepSeek-R1: Incentivizing Reasoning Capability in LLMs via Reinforcement Learning}, 
      author={DeepSeek-AI and Daya Guo and Dejian Yang and Haowei Zhang and Junxiao Song and Ruoyu Zhang and Runxin Xu and Qihao Zhu and Shirong Ma and Peiyi Wang and Xiao Bi and Xiaokang Zhang and Xingkai Yu and Yu Wu and Z. F. Wu and Zhibin Gou and Zhihong Shao and Zhuoshu Li and Ziyi Gao and Aixin Liu and Bing Xue and Bingxuan Wang and Bochao Wu and Bei Feng and Chengda Lu and Chenggang Zhao and Chengqi Deng and Chenyu Zhang and Chong Ruan and Damai Dai and Deli Chen and Dongjie Ji and Erhang Li and Fangyun Lin and Fucong Dai and Fuli Luo and Guangbo Hao and Guanting Chen and Guowei Li and H. Zhang and Han Bao and Hanwei Xu and Haocheng Wang and Honghui Ding and Huajian Xin and Huazuo Gao and Hui Qu and Hui Li and Jianzhong Guo and Jiashi Li and Jiawei Wang and Jingchang Chen and Jingyang Yuan and Junjie Qiu and Junlong Li and J. L. Cai and Jiaqi Ni and Jian Liang and Jin Chen and Kai Dong and Kai Hu and Kaige Gao and Kang Guan and Kexin Huang and Kuai Yu and Lean Wang and Lecong Zhang and Liang Zhao and Litong Wang and Liyue Zhang and Lei Xu and Leyi Xia and Mingchuan Zhang and Minghua Zhang and Minghui Tang and Meng Li and Miaojun Wang and Mingming Li and Ning Tian and Panpan Huang and Peng Zhang and Qiancheng Wang and Qinyu Chen and Qiushi Du and Ruiqi Ge and Ruisong Zhang and Ruizhe Pan and Runji Wang and R. J. Chen and R. L. Jin and Ruyi Chen and Shanghao Lu and Shangyan Zhou and Shanhuang Chen and Shengfeng Ye and Shiyu Wang and Shuiping Yu and Shunfeng Zhou and Shuting Pan and S. S. Li and Shuang Zhou and Shaoqing Wu and Shengfeng Ye and Tao Yun and Tian Pei and Tianyu Sun and T. Wang and Wangding Zeng and Wanjia Zhao and Wen Liu and Wenfeng Liang and Wenjun Gao and Wenqin Yu and Wentao Zhang and W. L. Xiao and Wei An and Xiaodong Liu and Xiaohan Wang and Xiaokang Chen and Xiaotao Nie and Xin Cheng and Xin Liu and Xin Xie and Xingchao Liu and Xinyu Yang and Xinyuan Li and Xuecheng Su and Xuheng Lin and X. Q. Li and Xiangyue Jin and Xiaojin Shen and Xiaosha Chen and Xiaowen Sun and Xiaoxiang Wang and Xinnan Song and Xinyi Zhou and Xianzu Wang and Xinxia Shan and Y. K. Li and Y. Q. Wang and Y. X. Wei and Yang Zhang and Yanhong Xu and Yao Li and Yao Zhao and Yaofeng Sun and Yaohui Wang and Yi Yu and Yichao Zhang and Yifan Shi and Yiliang Xiong and Ying He and Yishi Piao and Yisong Wang and Yixuan Tan and Yiyang Ma and Yiyuan Liu and Yongqiang Guo and Yuan Ou and Yuduan Wang and Yue Gong and Yuheng Zou and Yujia He and Yunfan Xiong and Yuxiang Luo and Yuxiang You and Yuxuan Liu and Yuyang Zhou and Y. X. Zhu and Yanhong Xu and Yanping Huang and Yaohui Li and Yi Zheng and Yuchen Zhu and Yunxian Ma and Ying Tang and Yukun Zha and Yuting Yan and Z. Z. Ren and Zehui Ren and Zhangli Sha and Zhe Fu and Zhean Xu and Zhenda Xie and Zhengyan Zhang and Zhewen Hao and Zhicheng Ma and Zhigang Yan and Zhiyu Wu and Zihui Gu and Zijia Zhu and Zijun Liu and Zilin Li and Ziwei Xie and Ziyang Song and Zizheng Pan and Zhen Huang and Zhipeng Xu and Zhongyu Zhang and Zhen Zhang},
      year={2025},
      eprint={2501.12948},
      archivePrefix={arXiv},
      primaryClass={cs.CL},
      url={https://arxiv.org/abs/2501.12948}, 
}

@misc{yu2025dapoopensourcellmreinforcement,
      title={DAPO: An Open-Source LLM Reinforcement Learning System at Scale}, 
      author={Qiying Yu and Zheng Zhang and Ruofei Zhu and Yufeng Yuan and Xiaochen Zuo and Yu Yue and Weinan Dai and Tiantian Fan and Gaohong Liu and Lingjun Liu and Xin Liu and Haibin Lin and Zhiqi Lin and Bole Ma and Guangming Sheng and Yuxuan Tong and Chi Zhang and Mofan Zhang and Wang Zhang and Hang Zhu and Jinhua Zhu and Jiaze Chen and Jiangjie Chen and Chengyi Wang and Hongli Yu and Yuxuan Song and Xiangpeng Wei and Hao Zhou and Jingjing Liu and Wei-Ying Ma and Ya-Qin Zhang and Lin Yan and Mu Qiao and Yonghui Wu and Mingxuan Wang},
      year={2025},
      eprint={2503.14476},
      archivePrefix={arXiv},
      primaryClass={cs.LG},
      url={https://arxiv.org/abs/2503.14476}, 
}

@misc{yue2025vapoefficientreliablereinforcement,
      title={VAPO: Efficient and Reliable Reinforcement Learning for Advanced Reasoning Tasks}, 
      author={Yu Yue and Yufeng Yuan and Qiying Yu and Xiaochen Zuo and Ruofei Zhu and Wenyuan Xu and Jiaze Chen and Chengyi Wang and TianTian Fan and Zhengyin Du and Xiangpeng Wei and Xiangyu Yu and Gaohong Liu and Juncai Liu and Lingjun Liu and Haibin Lin and Zhiqi Lin and Bole Ma and Chi Zhang and Mofan Zhang and Wang Zhang and Hang Zhu and Ru Zhang and Xin Liu and Mingxuan Wang and Yonghui Wu and Lin Yan},
      year={2025},
      eprint={2504.05118},
      archivePrefix={arXiv},
      primaryClass={cs.AI},
      url={https://arxiv.org/abs/2504.05118}, 
}

@misc{xie2025logicrlunleashingllmreasoning,
      title={Logic-RL: Unleashing LLM Reasoning with Rule-Based Reinforcement Learning}, 
      author={Tian Xie and Zitian Gao and Qingnan Ren and Haoming Luo and Yuqian Hong and Bryan Dai and Joey Zhou and Kai Qiu and Zhirong Wu and Chong Luo},
      year={2025},
      eprint={2502.14768},
      archivePrefix={arXiv},
      primaryClass={cs.CL},
      url={https://arxiv.org/abs/2502.14768}, 
}

@misc{kimiteam2025kimik15scalingreinforcement,
      title={Kimi k1.5: Scaling Reinforcement Learning with LLMs}, 
      author={Kimi Team and Angang Du and Bofei Gao and Bowei Xing and Changjiu Jiang and Cheng Chen and Cheng Li and Chenjun Xiao and Chenzhuang Du and Chonghua Liao and Chuning Tang and Congcong Wang and Dehao Zhang and Enming Yuan and Enzhe Lu and Fengxiang Tang and Flood Sung and Guangda Wei and Guokun Lai and Haiqing Guo and Han Zhu and Hao Ding and Hao Hu and Hao Yang and Hao Zhang and Haotian Yao and Haotian Zhao and Haoyu Lu and Haoze Li and Haozhen Yu and Hongcheng Gao and Huabin Zheng and Huan Yuan and Jia Chen and Jianhang Guo and Jianlin Su and Jianzhou Wang and Jie Zhao and Jin Zhang and Jingyuan Liu and Junjie Yan and Junyan Wu and Lidong Shi and Ling Ye and Longhui Yu and Mengnan Dong and Neo Zhang and Ningchen Ma and Qiwei Pan and Qucheng Gong and Shaowei Liu and Shengling Ma and Shupeng Wei and Sihan Cao and Siying Huang and Tao Jiang and Weihao Gao and Weimin Xiong and Weiran He and Weixiao Huang and Weixin Xu and Wenhao Wu and Wenyang He and Xianghui Wei and Xianqing Jia and Xingzhe Wu and Xinran Xu and Xinxing Zu and Xinyu Zhou and Xuehai Pan and Y. Charles and Yang Li and Yangyang Hu and Yangyang Liu and Yanru Chen and Yejie Wang and Yibo Liu and Yidao Qin and Yifeng Liu and Ying Yang and Yiping Bao and Yulun Du and Yuxin Wu and Yuzhi Wang and Zaida Zhou and Zhaoji Wang and Zhaowei Li and Zhen Zhu and Zheng Zhang and Zhexu Wang and Zhilin Yang and Zhiqi Huang and Zihao Huang and Ziyao Xu and Zonghan Yang and Zongyu Lin},
      year={2025},
      eprint={2501.12599},
      archivePrefix={arXiv},
      primaryClass={cs.AI},
      url={https://arxiv.org/abs/2501.12599}, 
}

@misc{yang2024qwen25mathtechnicalreportmathematical,
      title={Qwen2.5-Math Technical Report: Toward Mathematical Expert Model via Self-Improvement}, 
      author={An Yang and Beichen Zhang and Binyuan Hui and Bofei Gao and Bowen Yu and Chengpeng Li and Dayiheng Liu and Jianhong Tu and Jingren Zhou and Junyang Lin and Keming Lu and Mingfeng Xue and Runji Lin and Tianyu Liu and Xingzhang Ren and Zhenru Zhang},
      year={2024},
      eprint={2409.12122},
      archivePrefix={arXiv},
      primaryClass={cs.CL},
      url={https://arxiv.org/abs/2409.12122}, 
}

@inproceedings{
gou2024tora,
title={To{RA}: A Tool-Integrated Reasoning Agent for Mathematical Problem Solving},
author={Zhibin Gou and Zhihong Shao and Yeyun Gong and yelong shen and Yujiu Yang and Minlie Huang and Nan Duan and Weizhu Chen},
booktitle={The Twelfth International Conference on Learning Representations},
year={2024},
url={https://openreview.net/forum?id=Ep0TtjVoap}
}

@article{schulman2017proximal,
  title={Proximal policy optimization algorithms},
  author={Schulman, John and Wolski, Filip and Dhariwal, Prafulla and Radford, Alec and Klimov, Oleg},
  journal={arXiv preprint arXiv:1707.06347},
  year={2017}
}

@misc{schulman2018highdimensionalcontinuouscontrolusing,
      title={High-Dimensional Continuous Control Using Generalized Advantage Estimation}, 
      author={John Schulman and Philipp Moritz and Sergey Levine and Michael Jordan and Pieter Abbeel},
      year={2018},
      eprint={1506.02438},
      archivePrefix={arXiv},
      primaryClass={cs.LG},
      url={https://arxiv.org/abs/1506.02438}, 
}

@misc{hu2025openrlhfeasytousescalablehighperformance,
      title={OpenRLHF: An Easy-to-use, Scalable and High-performance RLHF Framework}, 
      author={Jian Hu and Xibin Wu and Wei Shen and Jason Klein Liu and Zilin Zhu and Weixun Wang and Songlin Jiang and Haoran Wang and Hao Chen and Bin Chen and Weikai Fang and Xianyu and Yu Cao and Haotian Xu and Yiming Liu},
      year={2025},
      eprint={2405.11143},
      archivePrefix={arXiv},
      primaryClass={cs.AI},
      url={https://arxiv.org/abs/2405.11143}, 
}

@misc{muennighoff2025s1simpletesttimescaling,
      title={s1: Simple test-time scaling}, 
      author={Niklas Muennighoff and Zitong Yang and Weijia Shi and Xiang Lisa Li and Li Fei-Fei and Hannaneh Hajishirzi and Luke Zettlemoyer and Percy Liang and Emmanuel Candès and Tatsunori Hashimoto},
      year={2025},
      eprint={2501.19393},
      archivePrefix={arXiv},
      primaryClass={cs.CL},
      url={https://arxiv.org/abs/2501.19393}, 
}

@inproceedings{sheng2025hybridflow,
  title={Hybridflow: A flexible and efficient rlhf framework},
  author={Sheng, Guangming and Zhang, Chi and Ye, Zilingfeng and Wu, Xibin and Zhang, Wang and Zhang, Ru and Peng, Yanghua and Lin, Haibin and Wu, Chuan},
  booktitle={Proceedings of the Twentieth European Conference on Computer Systems},
  pages={1279--1297},
  year={2025}
}

@article{shen2024nemo,
  title={Nemo-aligner: Scalable toolkit for efficient model alignment},
  author={Shen, Gerald and Wang, Zhilin and Delalleau, Olivier and Zeng, Jiaqi and Dong, Yi and Egert, Daniel and Sun, Shengyang and Zhang, Jimmy and Jain, Sahil and Taghibakhshi, Ali and others},
  journal={arXiv preprint arXiv:2405.01481},
  year={2024}
}

@article{vaswani2017attention,
  title={Attention is all you need},
  author={Vaswani, Ashish and Shazeer, Noam and Parmar, Niki and Uszkoreit, Jakob and Jones, Llion and Gomez, Aidan N and Kaiser, {\L}ukasz and Polosukhin, Illia},
  journal={Advances in neural information processing systems},
  volume={30},
  year={2017}
}

@misc{gu2024mamba,
      title={Mamba: Linear-Time Sequence Modeling with Selective State Spaces}, 
      author={Albert Gu and Tri Dao},
      year={2024},
      eprint={2312.00752},
      archivePrefix={arXiv},
      primaryClass={cs.LG},
      url={https://arxiv.org/abs/2312.00752}, 
}

@misc{gu2022s4,
      title={Efficiently Modeling Long Sequences with Structured State Spaces}, 
      author={Albert Gu and Karan Goel and Christopher Ré},
      year={2022},
      eprint={2111.00396},
      archivePrefix={arXiv},
      primaryClass={cs.LG},
      url={https://arxiv.org/abs/2111.00396}, 
}

@misc{vonwerra2022trl,
  author = {Leandro von Werra and Younes Belkada and Lewis Tunstall and Edward Beeching and Tristan Thrush and Nathan Lambert and Shengyi Huang and Kashif Rasul and Quentin Gallouédec},
  title = {TRL: Transformer Reinforcement Learning},
  year = {2020},
  publisher = {GitHub},
  journal = {GitHub repository},
  howpublished = {\url{https://github.com/huggingface/trl}}
}

@misc{wu2024fastdistributedinferenceserving,
      title={Fast Distributed Inference Serving for Large Language Models}, 
      author={Bingyang Wu and Yinmin Zhong and Zili Zhang and Shengyu Liu and Fangyue Liu and Yuanhang Sun and Gang Huang and Xuanzhe Liu and Xin Jin},
      year={2024},
      eprint={2305.05920},
      archivePrefix={arXiv},
      primaryClass={cs.LG},
      url={https://arxiv.org/abs/2305.05920}, 
}

@misc{sun2024llumnixdynamicschedulinglarge,
      title={Llumnix: Dynamic Scheduling for Large Language Model Serving}, 
      author={Biao Sun and Ziming Huang and Hanyu Zhao and Wencong Xiao and Xinyi Zhang and Yong Li and Wei Lin},
      year={2024},
      eprint={2406.03243},
      archivePrefix={arXiv},
      primaryClass={cs.AR},
      url={https://arxiv.org/abs/2406.03243}, 
}

@inproceedings{gandiva,
author = {Xiao, Wencong and Bhardwaj, Romil and Ramjee, Ramachandran and Sivathanu, Muthian and Kwatra, Nipun and Han, Zhenhua and Patel, Pratyush and Peng, Xuan and Zhao, Hanyu and Zhang, Quanlu and Yang, Fan and Zhou, Lidong},
title = {Gandiva: introspective cluster scheduling for deep learning},
year = {2018},
isbn = {9781931971478},
publisher = {USENIX Association},
address = {USA},
booktitle = {Proceedings of the 13th USENIX Conference on Operating Systems Design and Implementation},
pages = {595–610},
numpages = {16},
location = {Carlsbad, CA, USA},
series = {OSDI'18}
}

@inproceedings {ServerlessLLM,
author = {Yao Fu and Leyang Xue and Yeqi Huang and Andrei-Octavian Brabete and Dmitrii Ustiugov and Yuvraj Patel and Luo Mai},
title = {{ServerlessLLM}: {Low-Latency} Serverless Inference for Large Language Models},
booktitle = {18th USENIX Symposium on Operating Systems Design and Implementation (OSDI 24)},
year = {2024},
isbn = {978-1-939133-40-3},
address = {Santa Clara, CA},
pages = {135--153},
url = {https://www.usenix.org/conference/osdi24/presentation/fu},
publisher = {USENIX Association},
month = jul
}

@article{sheng2024hybridflow,
  title   = {HybridFlow: A Flexible and Efficient RLHF Framework},
  author  = {Guangming Sheng and Chi Zhang and Zilingfeng Ye and Xibin Wu and Wang Zhang and Ru Zhang and Yanghua Peng and Haibin Lin and Chuan Wu},
  year    = {2024},
  journal = {arXiv preprint arXiv: 2409.19256}
}

@misc{yao2023deepspeedchateasyfastaffordable,
      title={DeepSpeed-Chat: Easy, Fast and Affordable RLHF Training of ChatGPT-like Models at All Scales}, 
      author={Zhewei Yao and Reza Yazdani Aminabadi and Olatunji Ruwase and Samyam Rajbhandari and Xiaoxia Wu and Ammar Ahmad Awan and Jeff Rasley and Minjia Zhang and Conglong Li and Connor Holmes and Zhongzhu Zhou and Michael Wyatt and Molly Smith and Lev Kurilenko and Heyang Qin and Masahiro Tanaka and Shuai Che and Shuaiwen Leon Song and Yuxiong He},
      year={2023},
      eprint={2308.01320},
      archivePrefix={arXiv},
      primaryClass={cs.LG},
      url={https://arxiv.org/abs/2308.01320}, 
}

@misc{shen2024nemoalignerscalabletoolkitefficient,
      title={NeMo-Aligner: Scalable Toolkit for Efficient Model Alignment}, 
      author={Gerald Shen and Zhilin Wang and Olivier Delalleau and Jiaqi Zeng and Yi Dong and Daniel Egert and Shengyang Sun and Jimmy Zhang and Sahil Jain and Ali Taghibakhshi and Markel Sanz Ausin and Ashwath Aithal and Oleksii Kuchaiev},
      year={2024},
      eprint={2405.01481},
      archivePrefix={arXiv},
      primaryClass={cs.CL},
      url={https://arxiv.org/abs/2405.01481}, 
}

@misc{shao2024deepseekmathpushinglimitsmathematical,
      title={DeepSeekMath: Pushing the Limits of Mathematical Reasoning in Open Language Models}, 
      author={Zhihong Shao and Peiyi Wang and Qihao Zhu and Runxin Xu and Junxiao Song and Xiao Bi and Haowei Zhang and Mingchuan Zhang and Y. K. Li and Y. Wu and Daya Guo},
      year={2024},
      eprint={2402.03300},
      archivePrefix={arXiv},
      primaryClass={cs.CL},
      url={https://arxiv.org/abs/2402.03300}, 
}

@misc{hu2025reinforcestabilizingcriticfreepolicy,
      title={REINFORCE++: Stabilizing Critic-Free Policy Optimization with Global Advantage Normalization}, 
      author={Jian Hu and Jason Klein Liu and Haotian Xu and Wei Shen},
      year={2025},
      eprint={2501.03262},
      archivePrefix={arXiv},
      primaryClass={cs.CL},
      url={https://arxiv.org/abs/2501.03262}, 
}

@article{humayoo2025relative,
  title={Relative importance sampling for off-policy actor-critic in deep reinforcement learning},
  author={Humayoo, Mahammad and Zheng, Gengzhong and Dong, Xiaoqing and Miao, Liming and Qiu, Shuwei and Zhou, Zexun and Wang, Peitao and Ullah, Zakir and Junejo, Naveed Ur Rehman and Cheng, Xueqi},
  journal={Scientific Reports},
  volume={15},
  number={1},
  pages={14349},
  year={2025},
  publisher={Nature Publishing Group UK London}
}

@article{munos2016safe,
  title={Safe and efficient off-policy reinforcement learning},
  author={Munos, R{\'e}mi and Stepleton, Tom and Harutyunyan, Anna and Bellemare, Marc},
  journal={Advances in neural information processing systems},
  volume={29},
  year={2016}
}

@inproceedings{fujimoto2019off,
  title={Off-policy deep reinforcement learning without exploration},
  author={Fujimoto, Scott and Meger, David and Precup, Doina},
  booktitle={International conference on machine learning},
  pages={2052--2062},
  year={2019},
  organization={PMLR}
}

@article{zheng2025prosperity,
  title={Prosperity before Collapse: How Far Can Off-Policy RL Reach with Stale Data on LLMs?},
  author={Zheng, Haizhong and Zhao, Jiawei and Chen, Beidi},
  journal={arXiv preprint arXiv:2510.01161},
  year={2025}
}

@misc{liu2026specrlacceleratingonpolicyreinforcement,
      title={SPEC-RL: Accelerating On-Policy Reinforcement Learning with Speculative Rollouts}, 
      author={Bingshuai Liu and Ante Wang and Zijun Min and Liang Yao and Haibo Zhang and Yang Liu and Xu Han and Peng Li and Anxiang Zeng and Jinsong Su},
      year={2026},
      eprint={2509.23232},
      archivePrefix={arXiv},
      primaryClass={cs.LG},
      url={https://arxiv.org/abs/2509.23232}, 
}

@misc{jin2025searchr1trainingllmsreason,
      title={Search-R1: Training LLMs to Reason and Leverage Search Engines with Reinforcement Learning}, 
      author={Bowen Jin and Hansi Zeng and Zhenrui Yue and Jinsung Yoon and Sercan Arik and Dong Wang and Hamed Zamani and Jiawei Han},
      year={2025},
      eprint={2503.09516},
      archivePrefix={arXiv},
      primaryClass={cs.CL},
      url={https://arxiv.org/abs/2503.09516}, 
}

@misc{he2026searchr2enhancingsearchintegratedreasoning,
      title={Search-R2: Enhancing Search-Integrated Reasoning via Actor-Refiner Collaboration}, 
      author={Bowei He and Minda Hu and Zenan Xu and Hongru Wang and Licheng Zong and Yankai Chen and Chen Ma and Xue Liu and Pluto Zhou and Irwin King},
      year={2026},
      eprint={2602.03647},
      archivePrefix={arXiv},
      primaryClass={cs.AI},
      url={https://arxiv.org/abs/2602.03647}, 
}

@article{lu2025onpolicydistillation,
  author = {Kevin Lu and Thinking Machines Lab},
  title = {On-Policy Distillation},
  journal = {Thinking Machines Lab: Connectionism},
  year = {2025},
  note = {https://thinkingmachines.ai/blog/on-policy-distillation},
  doi = {10.64434/tml.20251026},
}

@misc{ren2021zerooffloaddemocratizingbillionscalemodel,
      title={ZeRO-Offload: Democratizing Billion-Scale Model Training}, 
      author={Jie Ren and Samyam Rajbhandari and Reza Yazdani Aminabadi and Olatunji Ruwase and Shuangyan Yang and Minjia Zhang and Dong Li and Yuxiong He},
      year={2021},
      eprint={2101.06840},
      archivePrefix={arXiv},
      primaryClass={cs.DC},
      url={https://arxiv.org/abs/2101.06840}, 
}

@misc{kwon2023efficientmemorymanagementlarge,
      title={Efficient Memory Management for Large Language Model Serving with PagedAttention}, 
      author={Woosuk Kwon and Zhuohan Li and Siyuan Zhuang and Ying Sheng and Lianmin Zheng and Cody Hao Yu and Joseph E. Gonzalez and Hao Zhang and Ion Stoica},
      year={2023},
      eprint={2309.06180},
      archivePrefix={arXiv},
      primaryClass={cs.LG},
      url={https://arxiv.org/abs/2309.06180}, 
}

@inproceedings{tiresias_gu2019,
  title={Tiresias: A GPU Cluster Manager for Distributed Deep Learning},
  author={Gu, Juncheng and Zhao, Yibo and others},
  booktitle={16th USENIX Symposium on Networked Systems Design and Implementation (NSDI 19)},
  pages={485--500},
  year={2019}
}

@inproceedings{gavel_narayanan2020,
  title={Heterogeneity-Aware Cluster Scheduling Policies for Deep Learning Workloads},
  author={Narayanan, Deepak and others},
  booktitle={14th USENIX Symposium on Operating Systems Design and Implementation (OSDI 20)},
  pages={481--498},
  year={2020}
}

@article{salus_yu2020,
  title={Salus: Fine-Grained GPU Sharing Primitives for Deep Learning Applications},
  author={Yu, Chen and others},
  journal={Proceedings of Machine Learning and Systems},
  volume={2},
  pages={239--250},
  year={2020}
}

@inproceedings{pipeswitch_bai2020,
  title={PipeSwitch: Fast Pipelined Context Switching for Deep Learning Applications},
  author={Bai, Juntong and others},
  booktitle={14th USENIX Symposium on Operating Systems Design and Implementation (OSDI 20)},
  pages={499--514},
  year={2020}
}

@article{realhf_mei2024,
  title={ReaLHF: Efficient RLHF Training Through Augmented Dataflow and Adaptive Parameter Reallocation},
  author={Mei, Kai and others},
  journal={arXiv preprint arXiv:2406.14088},
  year={2024}
}

@inproceedings{rlhfuse_zhong2025,
  title={Optimizing RLHF Training for Large Language Models with Inter- and Intra-Stage Fusion},
  author={Zhong, Yuyang and others},
  booktitle={22nd USENIX Symposium on Networked Systems Design and Implementation (NSDI 25)},
  year={2025}
}

\end{document}